\documentclass{article} 
 
\twocolumn
 
\begin{document}
 
\title{Physics of Aberration rather than Special 
Relativity}
 
\author{Yong Gwan Yi}
 
\maketitle

\noindent{\bf Abstract}\quad A phenomenological 
explanation is presented for the physics of 
aberration, which is in contrast with special 
relativity physics. The effect of relativity is 
identified with an effect due to the velocity of 
observation being affected by the velocity of a 
moving particle. In contrast with the currently 
accepted view, it is demonstrated that the 
classical concepts of time and simultaneity are 
natural for describing relativistic phenomena.
  
\bigskip
 
 
\vspace{50pt}
 
\noindent{\bf 1 Introduction}
 
\bigskip
 
Einstein's theory of special relativity has become a 
commonplace in modern physics, as taken for granted 
as Newton's law of classical mechanics or the 
Maxwell equations of electromagnetism. However, it 
was resisted for many years because of the second 
postulate on which the theory is based. The second 
postulate, which states that the speed of light is 
independent of the motion of its source, destroys 
the concept of time as a universal variable 
independent of the spatial coordinates. It forces on 
us a radical rethinking of our ideas about time and 
space. Many attempts were made to invent theories 
that would explain all the observed facts without 
this assumption. Our changed concept of time is the 
result of its gradual establishment through 
experiments in violent controversy.
  
This work is another such attempt. In contrast with 
previous works, I tried to pick out an essential 
physical point in the relativistic formalism. 
Attention was focused on the Lorentz condition which 
led to the formulation of special relativity. In 
this attempt, I have come to see a physics behind 
the aberration of starlight. In this paper, I 
present a phenomenological explanation for the 
physics of aberration. This is in contrast with the 
relativistic explanation of special relativity 
physics. It begins by reasoning a physical origin of 
relativistic phenomena, leading to the relativistic 
form of equations on the basis of classical physics. 
There is no need to make an assumption. We need to 
rethink some of the established thought and review 
the understanding of special relativity physics.
  
\vspace{30pt}
 
\noindent{\bf 2 Ether drift}
 
\bigskip
 
The Michelson-Morley experiment was undertaken to 
investigate the possible existence of an ether drift 
[1]. In principle, it consisted merely of observing 
whether there was any shift of the fringes in the 
Michelson interferometer when the instrument was 
turned through an angle of 90$^{\circ}$. 
Observations showed that the shift is at most but a 
small fraction of the predicted value. The negative 
result was explained as demonstrating the absence of 
the ether drift. However, it could have been due to 
the experiment itself being incapable of 
demonstrating the ether drift.
 
Fizeau performed an experiment to determine whether 
the speed of light in a material medium is affected 
by motion of the medium relative to the source and 
observer. The experiment is much in the same way 
as the Rayleigh refractometer except the tubes 
containing water flowing rapidly between the source 
and observer. An alteration of the speed of light 
was observed in the Fizeau experiment, which was in 
reasonable agreement with the value given by 
Fresnel's dragging formula. In the Michelson-Morley 
experiment, it is assumed that the ether is in 
uniform motion through the source and observer. As 
viewed from the Fizeau experiment, the ether drift 
cannot be assumed in this arrangement. The 
circumstances are the same as for the Earth, 
whose motion cannot be defined without an 
extraterrestrial reference. Even if the 
Michelson-Morley experiment is performed in water 
flowing rapidly in one direction, the null result 
is expected since the velocity of the water flow 
cannot be defined in this arrangement. In the case 
of sound under the same circumstances, no change 
of pitch is to be expected as remarked by Rayleigh 
about Doppler's principle [2].
 
We should mention the Michelson-Morley experiment 
performed with an extraterrestrial light source. 
Apparently, the motion of the light source relative 
to the half-silvered mirror is ineffective in 
changing the interference pattern. As shown in the 
Michelson interferometer, only the motion of the 
half-silvered mirror relative to one of the other 
two mirrors can give rise to an effect on the 
interference fringes. It is clear that the point 
of splitting into two beams plays the role of an 
effective source in that interferometer. The 
experiment using sunlight differs from the original 
only by taste rather than coverage. 
  
\vspace{30pt}
 
\noindent{\bf 3 Twin paradox}
 
\bigskip
 
Lorentz obtained transformation equations by using 
a covariant condition which preserves the speed of 
light in all uniformly moving systems. Einstein 
showed that the transformation equations with the 
covariant condition require revision of the usual
concepts of time and simultaneity, leading to the 
result that a moving clock runs more slowly than a 
stationary clock. Such a concept of time gives rise 
to the twin paradox, however. In mechanics, it is 
impossible by means of any physical measurements to 
label a coordinate system as intrinsically 
``stationary'' or ``uniformly moving''; one can only 
infer that the two systems are moving relative to 
each other. According to this fundamental postulate, 
like velocity and distance, time must also be 
symmetric with respect to the two systems. This is 
what the twin paradox points out.
 
We consider the experiments performed to verify the 
phenomenon of time dilation. The mean lifetime of 
$\pi$-mesons was determined using the decay of 
$\pi$-mesons at rest in a scintillator [3]. In this 
method, the mean lifetime of $\pi$-mesons was 
determined by a direct measurement of the time 
required to decay. In order to investigate the 
phenomenon of time dilation, an attempt to measure 
the mean lifetime of a rapidly moving $\pi$-meson 
beam was undertaken [4]. An experiment of this 
nature was arranged to measure the attenuation in 
flight of a $\pi$-meson beam of known lifetime using 
a scintillation counter telescope of a variable 
length. The measured mean free path was divided by 
the mean velocity to get the mean lifetime. The mean 
lifetime thus obtained, when the Lorentz time 
dilation was taken into account, was in fair 
agreement with the data measured in the rest 
system of $\pi$-mesons. It is generally recognized 
that these experiments have verified the phenomenon 
of time dilation. 
 
However, those experiments have an ambiguous 
bearing on the phenomenon of time dilation. In the 
latter experiment, the relativistic correction was 
made directly in the mean lifetime, keeping the 
particle velocity intact. This is otherwise without 
example in high-energy physics, where the 
relativistic correction has been made in the form of 
four-vector velocity. 

The four-vector velocity is conceived in the 
context of time and space, leading to the 
formulation of special relativity. The space 
components are defined as the rate of change of 
the path of a particle with respect to its proper 
time, the time component being defined as that of 
a light. Such a definition is a result of confusion, 
however, unless by intention. The four-vector 
velocity cannot be defined by the Lorentz time 
dilation; they are alternative conceptually. In 
fact, in that definition has the path dilation been 
disregarded. The mean free path measured in the 
experiment is not the distance of its proper 
lifetime but that multiplied by the $\gamma$ factor. 
Once the Lorentz time dilation is taken into 
account, there is no room for the four-vector 
velocity formulation. This is what we observe. 
Either the time dilation or the four-velocity can 
be consistent with the experimental result. From 
the experiment it is evident that the time dilation 
and the four-velocity are alternative. To see the 
definite result, the mean lifetime of a rapidly 
moving $\pi$-meson beam must be determined by direct 
measurement in experiment. The mean lifetime thus 
obtained will be the same as the data measured in 
the rest system of $\pi$-mesons if the twin paradox 
is the correct argument. Such an experiment has 
never been done in the past. Nevertheless, we can 
infer the result from a comparison with astronomical 
observations.

A series of observations by a new technique between 
1968 and 1970 indicated that the components making 
up the nucleus of radio source 3C279 were in motion 
[5]. The activity, which occurs on a scale of 
milliseconds of arc, could not have been detected 
with the techniques available before the early 
1970s. Surprisingly, the speed of the components was 
estimated to be about ten times the speed of light. 
The mysterious phenomenon received scientific 
attention, immediately. Some other quasars such as 
3C273 also turned out to be superluminal sources. 
From direct observations of the distances travelled 
and the times required it is reported that their 
nuclei contain components apparently flying apart at 
speeds exceeding the speed of light. The concept of 
the speed of light as a limiting speed of material 
particles, which has been confirmed in physics, has 
been questioned in astronomy.
 
It seems that the $\pi$-meson experiment and the 
observation of superluminal motion are equivalent. 
The only difference would be in their explanations. 
In physical meaning, the observation of superluminal 
motion is equivalent to an experiment that has 
measured directly the mean lifetime of a rapidly 
moving $\pi$-meson beam. It is certain therefore 
without requiring an explicit experiment that the 
mean lifetime of a rapidly moving $\pi$-meson beam 
obtained by direct measurement is equivalent to the 
mean lifetime in the $\pi$-rest system. Their
equivalence leads us to the conclusion that a 
particle velocity itself appears dilated to the 
observer, keeping time intact. It is then only 
natural to predict an equal ageing of twins in 
relative motion, by which the twin paradox is 
resolved naturally. The Lorentz time dilation is 
nothing more than a merely mathematical relation. 
The phenomenon of time dilation is nothing but 
a physical misconception of it. As pointed out by 
the twin paradox, the concept of time dilation
violates the relativity of uniformly moving systems.

\vspace{30pt}

\noindent{\bf 4 Aberration of light}
 
\bigskip
 
The Bradley observation of the aberration of 
starlight seems to be even more important to modern 
physics than previously thought. This is because the 
aberration effect can physically be interpreted as 
expressing an equation which is in contrast with 
the Lorentz condition leading to the formulation of 
special relativity. I would like to show a physics 
behind the aberration which is in contrast with 
special relativity physics.
 
In 1727, Bradley discovered an apparent motion of 
star which he explained as due to the motion of the 
Earth in its orbit. This effect, known as 
aberration, is quite distinct from the well-known 
displacements of the nearer stars known as parallax. 
Bradley's explanation of this effect was that the 
apparent direction of the light reaching the Earth 
from a star is altered by the motion of the Earth 
during propagation. The reason for this is much the 
same as that involved when a little girl walking in 
the rain must tilt her umbrella forward to keep the 
rain off her feet.
 
Let the vector {\bf v} represent the velocity of the 
Earth relative to a system of coordinates fixed in 
the solar system, and {\bf c} that of the light 
relative to the solar system. Then the velocity of 
the light relative to the Earth has the direction of 
{\bf c}$'$, which is the vector difference between 
{\bf c} and {\bf v}. This is the direction in which 
the telescope must be pointed to observe the star 
image on the axis of the instrument. When the 
Earth's motion is perpendicular to the direction of 
the star, the relation $c'^2-v^2=c^2$ follows from 
the vector difference. If we set $c'=kc$, we see 
that the observation is performed at speed $c'$ 
greater than when the Earth is at rest. Keeping in 
mind that the speed of light can be a measure of 
speed, the altered speed of observation may give 
rise to the same effect as would be the case if the 
velocity scale were altered at the moment of 
observation. Accordingly, the velocity of the Earth 
is supposed to be $v'=kv$ in relation to the 
observation. Taking this velocity of the Earth, the 
``Bradley'' relation becomes $c'^2-v'^2=c^2$. The 
velocity scale can then be written in the closed 
form $k=1/(1-v^2/c^2)^{1/2}$. This is just the 
$\gamma$ factor in special relativity. As a 
result, the angle of aberration $\alpha$ is given by
\begin{equation}
\sin\alpha=\beta,\quad\cos\alpha=1/\gamma,
\quad\mbox{and}\quad\tan\alpha=\gamma\beta,
\end{equation}
where $\beta=v/c$. The appearance as the velocity 
scale shows that the $\gamma$ factor is of an optic 
nature at the speed of observation. This means that 
the relativistic effect is in nature an optical 
phenomenon.
 
After this consideration, mention may be made of 
the difference between the present interpretation 
and the relativistic explanation. In the present 
interpretation, the velocity of the Earth and the 
velocity of light relative to it are respectively 
assumed to be $\gamma v$ and $\gamma c$, while the 
velocity of light relative to the solar system at 
rest is $c$. If the distance to the solar system 
is $R$, the distance to the Earth is $\gamma R$. 
Regardless of whether the Earth is at rest or in 
motion, consequently, the time required for light 
to reach the Earth is $R/c$. In the relativistic 
explanation, the velocity of the Earth and the 
velocity of light relative to it are respectively 
$v$ and $c$, whereas the velocity of light relative 
to the solar system at rest is assumed to be $c/
\gamma$ in the Earth's frame [6]. The time required 
to reach the Earth is here $\gamma R/c$. Although 
explanations are different, the same relations are 
given for the angle of aberration. For the 
Michelson-Morley experiment, however, they are 
different. In contrast to the relativistic 
explanation, the null result is expected from the 
present interpretation.
 
Having revealed the hidden nature of the aberration 
of starlight, we are going to examine its effect on 
the equations of motion in Newtonian mechanics. 
From the vector difference between {\bf c}$'$ and 
{\bf v}$'$ for the velocity of light, a derivative 
with respect to time gives the equation of 
corresponding accelerations
\begin{equation}
\frac{d{\bf c}'}{dt}-\frac{d{\bf v}'}{dt}=
\frac{d{\bf c}}{dt}=0.
\end{equation}
The scalar product of the accelerations in this 
equation with the corresponding velocity vectors is 
written
\begin{equation}
c'\frac{dc'}{dt}-v'\frac{dv'}{dt}=0,
\quad\mbox{so}\quad 
c\frac{d(\gamma c)}{dt}-v\frac{d(\gamma v)}{dt}=0.
\end{equation}
Equation (3) can also be obtained by differentiating 
the Bradley relation $c'^2-v'^2=c^2$ with respect to 
time. The kinetic energy $T$ is defined in general 
to be such that the scalar product of the force and 
the velocity is the time rate of change of $T$. In 
comparing (3) with the definition of $T$, the 
relativistic expression for kinetic energy is seen 
to be $T=\gamma mc^2$ [7]. In the present 
discussion, the mass has been treated as a constant 
[8]. The Bradley relation $c'^2-v'^2=c^2$ can then 
be expressed in terms of kinetic energy and 
momentum, which is the covariant energy-momentum 
equation with $T^2/c^2-p^2=m^2c^2$. There is no 
difficulty in obtaining the relativistic form of 
energy and momentum equations along the physical 
line of thought in the framework of classical 
mechanics.              

Because the aberration effect is ascribed to a 
change in the velocity of observation due to the 
motion of an observer, it is thought that 
relativistic phenomena would appear due to the 
measurement velocity being affected by a particle 
velocity. It is just like a vector difference 
between velocities. This illustrates why 
relativistic phenomena appear more pronounced as the 
velocity of particles approaches the velocity of 
light. The idea becomes clear. Is the effect of 
relativity just an effect due to the velocity of 
measurement being affected by the velocity of a 
particle? Understood as such, special relativity 
physics is identified itself as denoting the branch 
of physics which takes into consideration even the 
measurement velocity as affected by the particle 
velocity. This makes clear why the velocity of light 
appears in the equations of motion of a material 
particle. In this regard, a particle speed as fast 
as or faster than light, apart from the possibility 
of existence, is unobservable because such a 
particle goes beyond the limit of observation.

We suppose that the Earth is uniformly moving with 
velocity {\bf v} with respect to the solar system. 
For simplicity, let the origins of the coordinates 
of the Earth and the solar system be coincident at 
time $t=0$, at which time the star emits a pulse of 
light. If this pulse of light reaches the solar 
system at a time $t$, the propagation paths of the 
light to the solar system and the Earth are 
respectively given by $R=ct$ and $R'=c't$. Let $x$ 
and $x'$ be the respective projections of $R$ and 
$R'$ along the direction of {\bf v}. By the 
Pythagoras theorem, then, the geometric figure of 
aberration gives us the expression
\begin{equation}
c^2t^2-x^2=c'^2t^2-x'^2.
\end{equation}
 
\begin{figure}
\unitlength=1mm
\begin{picture}(100,60)(0,0)
\thicklines
\put(35,50){\line(0,-1){40}}
\put(35,10){\line(-1,0){30}}
\put(35,50){\vector(-3,-4){30}}
\put(35,50){\vector(-1,-2){20}}
\put(30,52){\large\bf star}
\put(33, 5){\large\bf O}
\put(13, 5){\large\bf x}
\put(2, 5){\large\bf x$'$}
\put(2,15){\large\bf c$'$t}
\put(12,15){\large\bf ct}
\put(80,50){\line(0,-1){40}}
\put(80,10){\line(-1,0){30}}
\put(80,50){\vector(-3,-4){30}}
\put(80,50){\vector(-1,-2){20}}
\put(75,52){\large\bf star}
\put(78, 5){\large\bf O}
\put(58, 5){\large\bf x}
\put(48, 5){\large\bf x$'$}
\put(48,15){\large\bf ct$'$}
\put(57,15){\large\bf ct}
\end{picture}
\caption{The aberration effect and the Lorentz 
condition}
\end{figure}
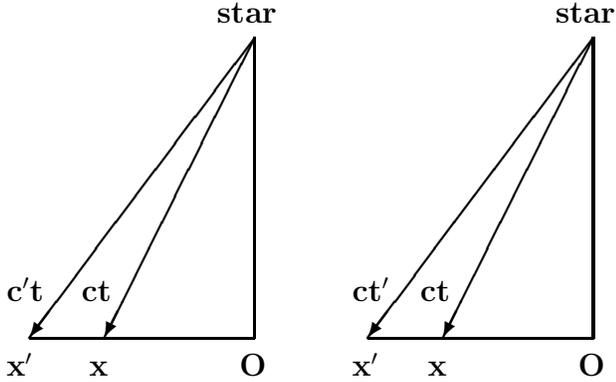
 
The general form of expression for aberration stands 
in contrast with the Lorentz condition which led to 
the transformation equations. It suggests taking 
$c't$ in place of $ct'$ as used in the Lorentz 
condition. They can be illustrated by the geometry 
of the Pythagoras theorem. In form, they correspond 
to an orthogonal transformation in a 
four-dimensional space consisting of the path of 
propagation of light and the three coordinates of 
space. It is important to notice their difference. 

The aberration of starlight shows the simultaneous 
arrival of light signals starting from the star at 
the two points $x$ and $x'$ in relative motion. The 
effect gives a physical interpretation for the 
four-dimensional space, which includes the 
observation in the description of motion. The 
Lorentz condition finds its explanation in a 
spreading spherical wave with time, which starts 
from the star and reaches the point $x$ at time $t$ 
and the point $x'$ at time $t'$. In the covariant 
form this gives the fourth coordinate as time. 
But it is given for the length of the path of 
propagation of light in terms of time. The Lorentz 
condition is a geometric relation. It has no 
physical bearing on a relative motion of the two 
points. With this very reason, the Lorentz 
transformation equations turn out to be the result 
of an ill-conceived marriage.
 
Seeing the Doppler effect, there is no doubt that
the velocity of light is not independent of the 
motion of its source. The invariance of the velocity 
of light in all uniformly moving systems, which 
plays so decisive a role in the Lorentz 
transformation, has an ambiguous bearing on the 
experimental facts. To be consistent with 
observation for the aberration of starlight, the 
Doppler shift, and the Michelson-Morley experiment, 
the second postulate should be replaced by the 
restricted, but more accurate, postulate that the 
velocity of light appears the same in all uniformly 
moving systems if and only if the source and the 
observer are both in a given system.
 
While a pulse of light propagates to the Earth, the 
motion of the Earth displaces its position: $x'=x-vt
$. In the same manner as derived the Lorentz 
transformation equations we can obtain an expression 
for the propagation path of starlight to the Earth. 
The aberration of starlight expressed in (4) can 
equally be solved to give 
\begin{equation}
c't=\gamma(ct-vx/c)\quad\mbox{or}\quad c'=\gamma 
c(1-\beta\cos\theta).
\end{equation}
Since the ratio between $x$ and $ct$ is the 
direction cosine of the propagation path of 
starlight with respect to {\bf v}, it can be 
expressed in the more familiar form of the Doppler 
shift formula. It is of interest to see that the 
aberration of starlight gives a general derivation 
of the relativistic formula for the Doppler shift. 
This leads us ultimately to consider the transverse 
Doppler shift as due to the aberration effect and 
thus as observed in the direction inclined at the 
angle of aberration toward the direction of motion 
of a moving source.
 
We can give a general derivation of the expression 
for the angle of aberration. As shown in the 
geometric figure, the ratio between the propagation 
path of starlight and the path of the Earth is a 
direction cosine. We obtain 
\begin{equation}
\cos\theta'=\frac{\cos\theta-\beta}{1-\beta\cos
\theta}\quad\mbox{from}\quad\frac{x'}{c't}=
\frac{\gamma(x-vt)}{\gamma(ct-vx/c)}.
\end{equation}
This is the same expression as given by considering 
the transformation of the phase of light wave, by 
Einstein [9]. 

It has been shown algebraically that two successive 
transformations with velocity parameters $\beta_1$ 
and $\beta_2$ are equivalent to a single Lorentz 
transformation of parameter $\beta=(\beta_1+
\beta_2)/(1+\beta_1\beta_2)$. This also follows from 
the ratio in (6), in consequence of the 
interpretation of $x/ct$ as the velocity parameter 
of a particle in the rest system and $x'/c't$ as the 
velocity parameter of observer in the laboratory. 
The formula for the addition of velocities comes 
from the inverse transformation equations. The 
inverse equations differ only by a change in the 
sign of {\bf v}. Note that the $\gamma$ factor is 
symmetric with respect to two systems in relative 
motion, the physics of relativity. It is misleading 
to introduce the relation $\Delta x'/\gamma=\Delta 
x$ as a basis for the Lorentz-FitzGerald contraction 
hypothesis from $\Delta x'=\gamma\Delta x$ given as 
a consequence of the Lorentz transformation 
equations.  
 
\vspace{30pt}
 
\noindent{\bf 5 Aberration of fields}
 
\bigskip
 
Newton's gravitational force is a static force. 
There is no notion of propagation, an action at a 
distance. In modern physics, it is required that a 
force be transmitted with a velocity. If the 
gravitational field propagates with the velocity of 
light instead of instantaneously, the gravitational 
field must suffer aberration, just as light does. 
It is then realized that the aberration of starlight 
expresses the aberration of the gravitational field 
of star.
 
Let {\bf R} be the radius vector from a star to the 
retarded position of the Earth. If the star is in a 
direction perpendicular to the motion of the Earth, 
the path of propagation of starlight to the Earth is 
given by $R/\cos\alpha=\gamma R$. The gravitational 
potential of the star can then be written as 
\begin{equation}
\biggl[\frac{GM}{R}\biggr]_{t-R/c} \mbox{and}\quad
\biggl[\frac{GM}{\gamma R}\biggr]_t.
\end{equation}
We may infer this form of gravitational potential 
from the aberration of starlight. It shows that the 
gravitational potential at the point of observation 
at time $t$ is determined by the state of motion of 
the Earth at the retarded time $t-R/c$, for which 
the time of propagation of light from the star to 
the observation point just coincides with $R/c$. 
 
We can extend this to the case where the star is not 
in a direction perpendicular to the motion of the 
Earth. The propagation path of starlight to the 
Earth is then given by (5) as $R'=\gamma R(1-\beta
\cdot {\bf n})$, where {\bf n} is a unit vector in 
the direction of {\bf R}. The gravitational 
potential can thus be written in the general form 
\begin{equation}
\frac{GM}{\gamma R(1-\beta\cdot{\bf n})}.
\end{equation}
If we define the gravitational field by the gradient 
of potential, then we obtain from the gravitational 
potential the expression 
\begin{equation}
\frac{GM}{\gamma^2R^2(1-\beta\cdot{\bf n})^2}
({\bf n}-{\bf \beta}),
\end{equation}
where we have used $\nabla R(1-\beta\cdot{\bf n}) 
={\bf n}-{\bf \beta}$ [10].
 
It is thought possible to express in a covariant 
form the aberration effect on the gravitational 
field. The gravitational field acting on the Earth 
is different in direction and magnitude from that 
when the Earth is at rest. In the geometric figure 
the difference is shown to be an acceleration that 
the moving Earth has during the propagation. The 
spatial variation in propagation of the 
gravitational field may be expressed in the form
\begin{equation}
\biggl[\frac{GM}{R^2}{\bf n} \biggr]_{t-R/c}
\Rightarrow\biggl[\frac{GM}{\gamma^2R^2
(1-\beta\cdot{\bf n})^2}({\bf n}-{\bf \beta})+
\frac{d(\gamma{\bf v})}{dt}\biggr]_t.
\end{equation}
This equation shows that the gravitational field 
acting on a moving system must be balanced by an 
acceleration the system would have during 
propagation. Total gravitational effects observed at 
a moving system will thus be the same, regardless of 
how fast it moves. This makes the gravitational 
field invariant in the covariant form. 

Following the same line of reasoning, the Coulomb 
potential produced by a moving electron can be 
expressed in the form of (8) by replacing the 
gravitational charge $GM$ by the electronic charge 
$e$. The Coulomb field thus obtained is in formal 
agreement with the electric field of an electron in 
uniform motion in electrodynamics. We can make a 
comparison with the Li\'{e}nard-Wiechert potential 
in terms of the retarded and present times:
\begin{equation}
\biggl[\frac{e}{R(1-\beta\cdot{\bf n})}\biggr]_{t-R/
c},\quad\biggl[\frac{e}{\gamma R(1-\beta\cdot{\bf 
n})}\biggr]_t.
\end{equation}
Since the relation of the retarded position to the 
present position of a moving electron is not, in 
general, known, the Li\'{e}nard-Wiechert potential 
ordinarily permits only the evaluation of the field 
in terms of retarded position and velocity of the 
electron. In the present approach, the unknown 
effect occurring during the propagation is assumed 
to be an aberration effect on the field attributed 
to its finite propagation velocity. As applied to a 
moving source of light, the aberration effect on the 
propagation of light to the observer yields an 
expression equal to the relativistic formula for the 
Doppler shift. This furnishes support for that
assumption. The unknown effect occurring during the 
propagation would be the aberration of the Coulomb 
field produced by a moving electron.
 
The electric field of a moving electron divides 
itself into a velocity field and an acceleration 
field [11]. In the present approach, the Coulomb 
potential alone induces the velocity field. Thus to 
make this approach agree with the electric field of 
a moving electron, the vector potential should be 
deduced solely from the acceleration field. On the 
assumption that the $\gamma$ factor is cancelled 
out by the relativistic correction to velocity, 
this deductive reasoning leads to the following 
expressions for the vector potential:
\begin{equation}
\frac{e}{c} \biggl[\frac{\bf v}{R(1-\beta\cdot{\bf 
n})}\biggr]_{t-R/c},\quad \frac{e}{c}
\biggl[\frac{{\bf v}-({\bf v}\cdot{\bf n}){\bf n}}
{R(1-\beta\cdot{\bf n})}\biggr]_t.
\end{equation}
This shows that the vector potential is evaluated 
by the component of velocity perpendicular to 
{\bf n}. When we view the vector potential in this 
way, we realize that the component of velocity 
parallel to {\bf n} has been incorporated in the 
velocity of field propagation. This makes it 
reasonable to expect the form of (12). Actually, 
it is true that the velocity of source appears 
as perpendicular to {\bf n} by the velocity of 
propagation relative to the velocity of source. 

The Li\'{e}nard-Wiechert potentials are to be 
evaluated at the retarded time. For derivatives, 
thus, we make use of transformations obtained by
differentiating $R=c(t-t_0)$:
\begin{equation}
\frac{\partial}{\partial t}=\frac 1{(1-{\bf\beta
\cdot n})}\frac{\partial}{\partial t_0},\quad 
\nabla=\nabla_R-\frac{\bf n}{c(1-{\bf 
\beta\cdot n})}\frac{\partial}{\partial t_0}. 
\end{equation}
When viewed from the present point, however, the 
aberration effect should be taken into 
consideration. In passing, we remark that the effect 
requires the vector potential to be transverse, 
satisfying the radiation gauge. In addition, the 
effect requires to evaluate the vector potential 
with respect to the path of propagation, $c'dt$ in 
place of $cdt$. In the radiation gauge, then, the 
electric field is given by
\begin{equation}
{\bf E}=-\frac 1c\frac{\partial{\bf A}}{\partial t}
\Rightarrow\frac ec\frac{{\bf n\times(v\times n)}}
{\gamma R^2(1-{\bf \beta\cdot n})^2}
+\frac {e}{c^2}\frac{\bf n\times\{(n-v)\times\dot{v}
\}}{\gamma R(1-{\bf \beta\cdot n})^3}.
\end{equation}
The first term is a result of differentiating $R$ 
by noting here $R=ct$. The second term is in 
agreement with the acceleration field except the 
$\gamma$ factor. As shown by (14), in form, the time 
derivative is equivalent to the differential 
operator. In the intuitive form, therefore, the 
magnetic induction may be evaluated in terms of the 
electric field:
\begin{equation}
{\bf B}=\nabla\times{\bf A}=-\frac{\bf n}{c}
\frac{\partial}{\partial t}\times{\bf A}=
{\bf n}\times{\bf E}.
\end{equation}
The aberration effect on the potential fields lends 
itself to incorporation in the classical theory of 
radiation.

We now consider the motion of an electron 
in a uniform magnetic field {\bf H}. If the electron 
has no velocity component along the field, it moves 
along a circle in the plane perpendicular to the 
field. The electron moving in the field satisfies 
the equation 
\begin{equation}
mv^2{\bf r}/r^2=e{\bf v}/c\times{\bf H}. 
\end{equation}
There would be an aberration of uniform magnetic 
field because of its finite propagation velocity. 
The physics of the situation is reminiscent of the 
aberration of starlight, where the field replaces 
starlight and the electron replaces the Earth in 
its orbit. The angle between {\bf v} and {\bf H} 
must be $\pi/2-\alpha$, instead of being $\pi/2$. 
The equation is written 
\begin{equation}
mv^2/r=(evH/c)\sin(\pi/2-\alpha).
\end{equation}
From the relation in (1), we find the magnetic 
frequency to be $eH/\gamma mc$. We can find a 
complete derivation of the relation for the magnetic 
frequency from the point of view of aberration. The 
$\gamma$ factor must be the aberration effect. 
 
Insight into the relativistic velocity of an 
electron can be provided by considering the 
mechanism by which the velocity of an electron is 
determined. An electrostatic spectrograph to 
determine the velocity of an electron consists in 
balancing the magnetic and electric deflections 
against each other [12]. The electron moving in a 
uniform magnetic field {\bf H}, perpendicularly to 
{\bf H}, describes a circular path of radius $R_H$:
\begin{equation}
mv^2{\bf R}_H/R_H^2=e{\bf v}/c\times{\bf H}.
\end{equation}
If this electron moves in a radial electric field 
{\bf E}, it can describe a circular path of radius 
$R_E$ given by 
\begin{equation}
mv^2{\bf R}_E/R_E^2=e{\bf E}.
\end{equation}
The equation of motion for the electron moving in 
the fields {\bf H} and {\bf E} applied 
simultaneously is then given by balancing the 
centrifugal force arising from the magnetic 
deflection against the centrifugal force due to the 
electric deflection, by
\begin{equation}
e{\bf E}R_E=e{\bf v}/c\times{\bf H}R_H.
\end{equation}
Taking into account the aberration occurring in the 
form of the vector difference between {\bf v} and 
{\bf H}, the angle between {\bf v} and {\bf H} is 
tilted at an angle of aberration toward the 
direction of motion of the moving electron. Thus, 
\begin{equation}
cER_E=vHR_H\sin(\pi/2-\alpha).
\end{equation}
The velocity of the electron is found to be 
$\gamma cER_E/HR_H$, where $\beta= ER_E/HR_H$. 
In this regard, $cER_E/HR_H$ is seen to be the 
intrinsic velocity the electron would have if the 
velocity of propagation of the fields were infinite, 
thereby not suffering aberration. This elucidates 
why a particle velocity itself appears dilated to 
the observer. The speed of high-energy particles of 
$\gamma v$ can easily be superluminal 
phenomenologically. It should be noted that the 
apparent speed of high-energy particles is ascribed 
to the aberration of uniform magnetic field. The 
relativistic velocity is identified with the 
apparent velocity of which the $\gamma$ factor 
arises out of the effect of aberration.

\vspace{30pt}
 
\noindent{\bf 6 Covariant Maxwell equations}
 
\bigskip 
 
We consider the electromagnetic fields seen by an 
observer in the system $S$ when a point charge $q$ 
moves by in a straightline path along the $x$ 
direction with a velocity {\bf v}. Let $S'$ be the 
moving coordinate system of $q$. The charge is at 
rest in this system. But when viewed from the system 
$S$, the charge represents a current ${\bf J}=q{\bf 
v}$ in the $x$ direction. The electromagnetic fields 
are then related through Amp\`{e}re's law:
\begin{equation}
\biggl[\nabla\times{\bf B} = \frac{4\pi}{c}{\bf J}
+\frac{1}{c}\frac{\partial{\bf E}}{\partial t}
\biggr]_S
=\ \biggl[\nabla\times{\bf B} = \frac{1}{c}
\frac{\partial{\bf E}}{\partial t}\biggr]_{S'}.
\end{equation}
 
Amp\`{e}re's law keeps its form invariant with 
respect to the two systems in relative motion. 
They are related at the same time, and so are 
the equations: $t=t'$. In the covariant form, 
nonetheless, it is instructive to write the 
equations of transformation between $S$ and $S'$ in 
terms of $t$ and $t'$. For that purpose, instead of 
using $ct$ and $c't$, we use here the Lorentz 
transformation equations. 

Let us apply to the equation the Lorentz 
transformation of coordinates with $[\gamma(ct-
\beta x),\gamma(x-vt), y, z]_S=[ct, x, y, z]_{S'}$. 
The $y$ and $z$ components are homogeneous 
equations. The transformation of these components 
is straightforward. The $x$ component is an 
inhomogeneous equation. Its transformation does not 
seem to be so.

By Coulomb's law $\nabla\cdot{\bf E}=4\pi q$, 
the equation can be written as 
\begin{equation}
\frac{\partial B_z}{\partial y}-\frac{\partial
B_y}{\partial z}=\frac{v}{c}(\nabla\cdot{\bf E})+
\frac{1}{c}\frac{\partial E_x}{\partial t}.
\end{equation}
If we multiply the $\gamma$ factor and use the 
inverse equations, we can transform the equation
into the form
\begin{equation}
\frac{\partial}{\partial y'}\biggl\{\gamma\biggl
(B_z-\frac{v}{c}E_y\biggr)\biggr\}-\frac{\partial}
{\partial z'}\biggl\{\gamma\biggl(B_y+\frac{v}{c}
E_z\biggr)\biggr\}\nonumber\\=\frac{1}{c}
\frac{\partial E_x}{\partial t'}.
\end{equation}
 
We may start with Faraday's law. In exactly the 
same manner, we use the relation $\nabla\cdot{\bf B} 
=0$ to obtain the equations of transformation. This 
completes the transformation of electromagnetic 
fields from the Maxwell equations.
 
\vspace{30pt}
 
\noindent{\bf 7 Concluding remarks}
 
\bigskip
 
We are taught special relativity in such a way that 
the phenomenon of time dilation is daily verified 
in high-energy physics laboratories. But the 
verification is not so explicit; one can only infer 
the lifetime dilation from the mean free path 
for the $\pi$-meson decay measured in the 
experiment. Nor are we unanimous in accepting or 
interpreting the concept of time dilation. The 
superluminal motion is by no means mysterious. 
The astronomical observation has shown us that a 
particle velocity itself appears dilated to the 
observer phenomenologically. Had the time been 
measured directly, the $\pi$-meson experiment would 
have shown essentially the same. Not only the 
experiment but the theory is incomplete. The 
aberration of uniform magnetic field has been 
overlooking in physics. The effect of aberration 
gives rise to the $\gamma$ factor of velocity, 
which disproves the phenomenon of time dilation. 
For the relativistic mass, likewise, the aberration 
effect disproves the experimental result. The effect 
of relativity is due to the $\gamma$ factor of 
velocity arising out of aberration. 

From special relativity we learn that the equations 
of motion should be covariant in the mathematical 
structure of time and space. By identical treatment 
of time and space, as Minkowski addressed [13], the 
forms in which the equations of motion are displayed 
gain in covariance. The Lorentz transformation 
equations were obtained by applying a covariant 
condition to two systems in relative motion. 
In the relative motion of two systems, however, it 
is assumed that time is the same in both systems. 
Two systems in relative motion cannot be covariant 
in time and space. The covariant condition can be 
satisfied by an equation for motion of a system or 
relative motion of two systems, providing a geometry 
in time and space for motion. But two systems in 
relative motion must not be confused with a relative 
motion of two systems. As noted by Sommerfeld [14], 
the fourth coordinate is not $t$ but $ct$. In the 
case of a moving source of light, furthermore, it is 
the velocity of light that appears dilated to the 
observer.
 
We can find in the effect of aberration a 
phenomenological explanation of special relativity 
physics. This reflects that the physical origin of 
relativistic phenomena lies in the aberration of 
starlight. The emphasis should be on how the physics 
of special relativity can be replaced in form and 
content by the physics of aberration. In contrast 
with special relativity, this leads us to an 
understanding of relativistic phenomena using 
a physical reasoning. It is demonstrated that the 
usual concepts of time and simultaneity are natural 
for describing relativistic phenomena. Einstein's 
argument is in essence a mathematical explanation 
based on the transformation equations. The resulting 
equations of Einstein's theory had been proved to 
be correct, contributing greatly to modern physics. 
However, the correct result does not always warrant 
the correctness of assumption. In the past 
controversy, the incorrect argument is not in 
opponents' minds but in Einstein's theory assuming 
the dilation of time scales. The concept of time 
dilation makes no sense physically; time is an 
independent variable and motion is relative to each 
other.

\vspace{30pt}
 
\noindent{\bf Appendix: Remark on the superluminal 
motion}
 
\bigskip
 
There has been a precision measurement of the 
neutrino velocity at 17 GeV with the OPERA detector 
at the underground Gran Sasso Laboratory [15]. The 
neutrino speed is measured by passing through about 
730 km of the Earth's crust from the CERN, showing 
values equivalent to the light speed within 
experimental errors. Intense debate on the 
experiment increases our interest in the OPERA 
result [16]. At much higher energy, the amazing 
result is compatible with earlier measurement of 
the neutrino velocity at 3 GeV from the Fermilab 
NuMI beam with the MINOS detector [17]. 
 
In the early 1970s, we were aware of a superluminal 
motion from the observation of radio source 3C279. 
Most astronomers could not believe the motion to be 
the case because the superluminal velocity cannot 
be accepted by the theory of special relativity. 
The current explanation given in astronomy must be 
reasonable, but it cannot be a physical explanation 
for the superluminal motion.

From a phenomenological point of view it is evident
that a particle velocity itself appears dilated to 
the observer by the $\gamma$ factor. The 
superluminal motion of jet in quasars must be 
such an apparent velocity. In fact, the intrinsic 
speed of the jet has been calculated by using the 
$\gamma$ factor required for the apparent velocity 
measured in the jet of 3C279 [18]. The derivation 
of apparent velocity is detailed in astronomy. 
But it is the aberration effect. It is due to 
the vector difference between the velocities of 
jet and light. A pulse of light emitted by the jet 
is propagated to us in the apparent direction. 
Like the velocity, we may deduce the intrinsic 
direction from the jet image. 
 
The neutrino velocity does not seem to be of the 
same character. The neutrino velocity has been 
determined with high accuracy through the 
measurement of the time of flight and the distance 
between the source of the neutrino beam at CERN 
and the OPERA detector at Gran Sasso Laboratory. 
The neutrino velocity cannot be an apparent 
velocity; the neutrino itself must be moving at 
such speed. Notice no effect of relativity in such 
measurement. Then the motion of neutrino is 
unobservable because the observation cannot catch 
up in speed with the neutrino. We can only observe 
the track of neutrino. This suggests their speed 
for why neutrinos could not be detected directly. 
Their negligible mass and neutral charge are 
technical reasons. In principle, we cannot apply 
the energy-momentum equation to the motion of 
neutrinos. This is because the four-momentum 
equation is given for the motion of a particle 
which is observable at the speed of light. 
Neutrino physics is a new physics beyond the 
observation and description of motion. 
 
\newpage
 
\noindent{\bf References}
 
\bigskip
 
\noindent[1] F. T. Jenkins and H. E. White, 
Fundamentals of Optics (McGraw-Hill, 1976) 4th ed., 
p. 416
 
\noindent[2] J. W. S. Rayleigh, The Theory of Sound
(Dover, 1945) vol. 2, p. 155
 
\noindent[3] M. Jakobson, A. Shulz, and J. 
Steinberger, Phys. Rev. {\bf 81} (1951) 894; C. E. 
Wiegand, Phys. Rev. {\bf 83} (1951) 1085
 
\noindent[4] R. P. Durbin, H. H. Loar, and W. W. 
Havens Jr, Phys. Rev. {\bf 88} (1952) 179

\noindent[5] Review article, ``Quasar's Jet; Faster 
Than Light?'' Science in Korea (1982) 24
 
\noindent[6] W. K. H. Panofsky and M. Phillips, 
Classical Electricity and Magnetism (Addison-Wesley, 
1962) 2nd ed., p. 303
 
\noindent[7] H. Goldstein, Classical Mechanics
(Addison-Wesley, 1950), p. 202
 
\noindent[8] C. G. Adler, Am. J. Phys. {\bf 55} 
(1987) 739; L. B. Okun, Phys. Today (June 1989) 31
 
\noindent[9] H. A. Lorentz, A. Einstein, H. 
Minkowski, and H. Weyl, The Principle of Relativity 
(Dover, 1952), p. 56

\noindent[10] Reference 6, p. 356

\noindent[11] J. D. Jackson, Classical 
Electrodynamics (John Wiley \& Sons, 1975) 2nd ed., 
p. 657
 
\noindent[12] M. M. Rogers, A. W. McReynolds, and 
F. T. Rogers Jr, Phys. Rev. {\bf 57} (1940) 379
 
\noindent[13] Reference 9, p. 75
 
\noindent[14] Reference 9, p. 92
 
\noindent[15] OPERA Collaboration, arXiv:1109.4897 
(2011)
 
\noindent[16] A. G. Cohen and S. L. Glashow, 
arXiv:1109.6562 (2011); ICARUS Collaboration, 
arXiv:1110.3763 (2011)
 
\noindent[17] MINOS Collaboration, arXiv:0706.0437 
(2007)

\noindent[18] B. G. Piner et al., 
arXiv:astro-ph/0301333 (2003)

\end{document}